\documentstyle[twocolumn,twoside,floats,prb,aps,psfig]{revtex}

\begin{document}
\newcommand{\beq}{\begin{equation}}
\newcommand{\eeq}{\end{equation}}
\newcommand{\degree}{$^{\rm\circ}$\ }

\title{A Minimal Model of B-DNA}

\author{Alexey K. Mazur}

\address{Laboratoire de Biochimie Th\'eorique, CNRS UPR9080\\
Institut de Biologie Physico-Chimique\\
13, rue Pierre et Marie Curie, Paris,75005, France.\\
FAX:(33-1) 43.29.56.45. Email: alexey@ibpc.fr}

\date{\today}
\maketitle

\begin{abstract}
Recently it has been found that stable and accurate
molecular dynamics (MD) of B-DNA duplexes can be obtained in
relatively inexpensive computational conditions
with the bulk solvent represented implicitly, but
the minor groove filled with explicit water
(J. Am. Chem. Soc. {\bf 1998}, 120, 10928). The present paper
further explores these simulation conditions in order
to understand the main factors responsible for the observed
surprisingly good agreement with experimental data.
It appears that in the case of the {\em Eco}RI dodecamer
certain sequence specific hydration patterns
in the minor groove earlier known from experimental data
are formed spontaneously in the course of MD simulations.
The effect is reliably reproduced in several independent
computational experiments in different simulation conditions.
With all major groove water removed, closely similar
results are obtained, with even better reproducibility.
On the other hand, without explicit hydration, metastable
dynamics around a B-DNA like state can be obtained which,
however, only poorly compares with experimental data.
It appears, therefore, that a right-handed DNA helix with
explicitly hydrated minor groove is a minimal model system
where the experimental properties of B-DNA can be reproduced
with non-trivial sequence-dependent effects. Its small size
makes possible virtually exhaustive sampling,
which is a major advantage with respect to alternative approaches.
An appendix is included with a correction to the implicit
leapfrog integrator used in internal coordinate MD.

\end{abstract}
\section{Introduction}
Understanding of the detailed molecular mechanisms of conformational
dynamics of the double helical DNA is an important goal on the long
way to control of genom functions. Computer simulations are the main
theoretical instrument in such studies. An important recent progress
in the methodology \cite{AMBER94:,MacKerell:95,Darden:93} has made
possible molecular dynamics (MD) simulations of short DNA fragments
in realistic water environment with explicit counterions
(comprehensive surveys of the literature can be found in the recent reviews
\cite{Ravishanker:97,Auffinger:98}). In should be noted, however,
that although in future these methods will
probably be able to describe in full detail
all interactions involved in the numerous biological functions of DNA,
at present their capabilities are very limited. There are many
important domains where they can hardly be applied, in the foreseeable
future, for instance, dynamics of long linear DNA fragments or
plasmids. Even for small oligomers these methods
require supercomputer resources because the solute
molecule must be placed in a large enough water box to accommodate all
neutralizing counterions without exceeding reasonable levels of
effective DNA and salt concentrations. This makes any methodological
questions difficult, especially those concerning statistical
reproducibility of results because, until now, almost all reports on
the subject described observations made for single trajectories.

The foregoing arguments explain why it is necessary to find
alternative conditions for MD simulations of B-DNA which would be less
computationally demanding, but yet acceptably accurate. The utility of
such DNA model would be twofold. On the one hand, it can give a
unique opportunity to probe the properties of large systems of real
biological importance where more rigorous approaches would be
prohibitively expensive. On the other hand, for smaller molecules they
can provide for an exhaustive sampling of the configurational space,
which is a necessary prerequisite of a more systematic studies of detailed
mechanisms involved in the conformational dynamics of double helical
DNA. We have recently found that surprisingly stable dynamics of B-DNA
duplexes can be obtained with semi-explicit treatment of long-range
electrostatics effects. \cite{Mzjacs:98} Paradoxically, it appeared
that this simplistic approach results in B-DNA conformations that
are significantly closer to experimental data than ever before,
including the recent more rigorous and expensive calculations. This
relative success raised several intriguing questions concerning its
origin. On the one hand, a better agreement with experimental data can
always result from a fortunate cancelation of errors. On the other
hand, it is possible that the earlier reported {\em a priori} more
rigorous calculations failed to reveal the full accuracy of the recent
atom parameter sets, notably AMBER94, \cite{AMBER94:} because of
insufficient duration of trajectories or some
problem in calculation of long range electrostatic interactions
by the particle-mesh Ewald technique. \cite{Darden:93}
Both these possibilities should be checked, but the
former is certainly easier to analyze because this does not require
expensive computations.

In this paper we address some of the above questions by further
exploring the dynamic behavior of B-DNA duplexes with partial explicit
hydration shell. Our main objective is to justify what we call a
``minimal B-DNA'', that is the least computationally demanding
simulation conditions where the essential experimental features of
this molecule can be reliably reproduced. It is shown that a good
candidate is a B-DNA duplex with explicitly hydrated minor groove,
empty major groove, and the electrostatic effects of the bulk solvent
accounted for implicitly. We report also about relatively successful attempts
to obtain stable dynamics without explicit hydration, which, however,
give much worse agreement with experimental data. In contrast,
dynamics of the minimal model reliably reproduces certain well-known sequence
dependent effects, notably, modulation of the minor groove width and
spontaneous formation of specific ``spine'' hydration pattern observed
in many experimental structures.  At the same time, for two very
different sequences it converges to significantly different
conformations which, however, are both close to the canonical B-DNA
form.  These results appear relatively insensitive to variations of
parameters involved in the simplified treatment of long-range
electrostatic interactions.

\section{Methods and Simulation Protocols}
All new MD simulations reported here have been performed with
the internal coordinate molecular dynamics (ICMD) method.
\cite{Mzjbsd:89,Jain:93,Mzjcc:97,Mzprep:98,Mzjchp:99}
Here we employ a modified implicit leapfrog integrator detailed in the Appendix.
We consider dynamics of two different DNA duplexes with
sequences d(CGCGAATTCGCG)$_2$ and d(GCGCGCGCGCGC)$_2$.
Most of the calculations were made with the first molecule often
referred to in the literature as ``{\em Eco}RI dodecamer''.
The second dodecamer is used for comparison.
The {\em Eco}RI sequence was the first to
crystallize in the B-form of DNA \cite{Drew:81} and since then it
became perhaps the most studied DNA fragment both experimentally and
theoretically.
\cite{Levitt:83b,Rao:90,Srinvasan:90,Swaminathan:91,Miaskiewicz:93,Kumar:94,McConnell:94}
In the recent years it has been often used in benchmark tests of new
force fields and algorithms.
\cite{Mzjacs:98,York:95,Yang:96,MacKerell:97,Duan:97,Young:97,Young:97a,Cieplak:97,Sprous:98,Jayaram:98,Winger:98}
The DNA model was same as in our recent report,
\cite{Mzjacs:98} namely, all torsions were free as
well as bond angles centered at sugar atoms, while other bonds and
angles were fixed, and the bases held rigid. The dynamic properties of
this model, notably, the fastest motions and maximal possible time
steps have already been studied. \cite{Mzjacs:98} AMBER94
\cite{AMBER94:} force field and atom parameters were
used with TIP3P water \cite{TIP3P:} and no cut off schemes.

The solvent effects were taken into account by the
mixed strategy \cite{Mzjacs:98} which we continue to explore and
improve here. In this method, following to a long known approximate
approach, \cite{Tidor:83,Mazur:91,JUMNA:95}
long range effects are taken into account
implicitly by reducing phosphate charges and using a linear distance
dependent dielectric function $\varepsilon = r$. On the other hand,
the minor groove of the DNA molecule is filled up by explicit water
with the water cloud somewhat protruding in space. In the initial
hydration procedure the DNA molecule is first covered by a 5~\AA\
thick water shell by placing it in a water box and eliminating
overlapping and distant solvent molecules. To allow water molecules to
penetrate into the minor groove, rather small cut-off distances are
used as the overlap criteria, namely, 1.8 and 1.3~\AA\ for oxygen and
hydrogen water atoms, respectively. After that cylinder-like volumes
around each strand are built from spheres centered at phosphorus atoms
with radii of 12~\AA. All water molecules that appear outside the
intersection area of the two volumes are removed. The solvent
remaining is next relaxed by energy minimization first with the solute
held rigid and then with all degrees of freedom. This procedure gives
a partially hydrated duplex, with the minor groove completely filled
and a few solvent molecules in the major groove. In some calculations
we additionally cleaned the major grove from the remaining water after
the second minimization. The motivations for this additional step are
discussed in the text.

The standard heating and equilibration protocols have been slightly
modified compared with our previous studies. Dynamics was initiated
with zero solvent temperature by giving the solute a kinetic energy
corresponding to 300~K. During the initial short 2 ps run the
temperature was weakly coupled to 250~K by the Berendsen algorithm
\cite{Berendsen:84} with a relaxation time of 1 ps. During the next 5
ps the temperature coupling was switched off and at the end of the
whole 7 ps period the system temperature normally reached 200~K.
Starting from this state the production trajectory was computed with
the temperature coupled to 300~K with a relaxation time of 10 ps. This
algorithm provides for a softer start and helps to reduce dissociation
of hot water molecules during the early non-equilibrated phases of
dynamics. \cite{Mzjacs:98} For trajectories starting
from the canonical B-DNA form the
energy minimization was applied both before and after hydration.
Hydration of the Xray structure \cite{Drew:81} was performed with
all crystallographic water molecules kept in their places.

Normally we used time steps of 5 and 10 fs in heating and production
runs, respectively. The possibility and the necessary conditions for
this large time steps have been discussed elsewhere.
\cite{Mzjacs:98,Mzjcp:97,Mzjpc:98}
Duration of trajectories varied depending upon
specific conditions discussed in the text. In calculations with the
solvent shell a minimum duration of 5 ns was usually sufficient to
arrive at stable levels of average helical parameters as well as
rmsd's from the reference conformations. In production runs
conformations were saved with a 2.5 ps interval. Structures from the
last nanosecond were used for computing an average conformation
referred to as the final MD state of the corresponding trajectory. The
reference canonical A and B forms were constructed with NUCGEN utility
of AMBER \cite{AMBER:} or JUMNA program. \cite{JUMNA:95} Curves
\cite{Curves:} and ``Dials and Windows'' procedures \cite{CDW:} were
used for analysis of DNA conformations. XmMol program \cite{XmMol:}
was used for visual inspection and animation of trajectories.

\begin{table*}[t]\caption{\label{Ttjli} General Information about
Trajectories}
\begin{tabular}[t]{|ddddddd|}
Code & Sequence & From  & Water\tablenotemark[1] & Clean\tablenotemark[2]
& Q$_p$\tablenotemark[3] (aeu) & Duration (ns) \\\hline
TJA\tablenotemark[4]
    &CGCGAATTCGCG& B73\tablenotemark[5]   & 134  &   -   &   -0.5   &  5.0 \\
TJB\tablenotemark[4]
    &CGCGAATTCGCG& {\em Eco}RI\tablenotemark[6]
                                          & 114  &   -   &   -0.5   &  5.1 \\
TJC\tablenotemark[4]
    &CGCGAATTCGCG& B73                    & 134  &   -   &   -0.5   &  5.0 \\
TJD &CGCGAATTCGCG& Xray\tablenotemark[7]  & 131  &   -   &   -0.5   &  8.8 \\
TJE &CGCGAATTCGCG& B73                    &  -   &   -   &   -0.25  &  2.7 \\
TJF &CGCGAATTCGCG& B73                    &  -   &   -   &   -0.5   &  5.0 \\
TJG &CGCGAATTCGCG& B73                    &  -   &   -   &   -0.75  &  2.9 \\
TJH &CGCGAATTCGCG& Xray                   & 175  &   +   &   -0.25  &  5.0 \\
TJI &CGCGAATTCGCG& Xray                   & 175  &   +   &   -0.5   &  5.5 \\
TJJ &CGCGAATTCGCG& Xray                   & 175  &   +   &   -0.6   &  8.0 \\
TJK &CGCGAATTCGCG& Xray                   & 175  &   +   &   -0.75  &  6.9 \\
TJL &GCGCGCGCGCGC& B73                    & 166  &   +   &   -0.5   &  8.1 \\
\end{tabular}
\tablenotetext[1]{
The total number of explicit water molecules
left after the hydration procedures described in the text.}
\tablenotetext[2]{
Major groove water removed before equilibration.}
\tablenotetext[3]{
Reduced total phosphate charge.}
\tablenotetext[4]{
Trajectories reported and discussed in detail in Ref. \onlinecite{Mzjacs:98}
}
\tablenotetext[5]{
The canonical B-DNA form with Arnott B73 parameters. \cite{Arnott:72} }
\tablenotetext[6]{
The `kinked' duplex conformation found in the {\em Eco}RI endonuclease complex.
\cite{McClarin:86} (file PDE001 in Nucleic Acids Database \cite{NDB:}) }
\tablenotetext[7]{
The crystal conformation. \cite{Drew:81}
(file 1bna in Protein Database. \cite{PDB:}) }
\end{table*}

\section{Results and Discussion}
\subsection{General Outline of Simulations and Results}
\begin{figure}
\centerline{\psfig{figure=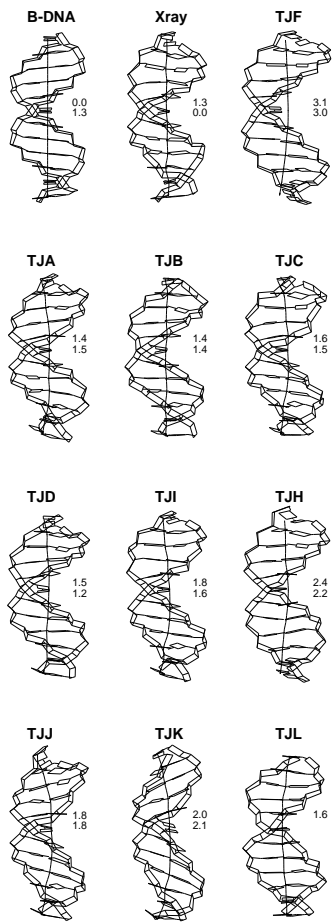,height=13cm,angle=0.,%
bbllx=300bp,bblly=80bp,bburx=560bp,bbury=720bp,clip=t}}
\caption{\label{Falstr}
The final states of representative trajectories in Table
\protect\ref{Ttjli}, together with the canonical B-form
and the Xray structure of the {\em Eco}RI dodecamer.
The optimal helical axis \protect\cite{Curves:} is shown by continuous
roughly vertical lines. On the right of each structure
the rmsd's from the canonical B-DNA and the Xray conformation
\protect\cite{Drew:81} are given.}
\end{figure}

\begin{figure}
\centerline{\psfig{figure=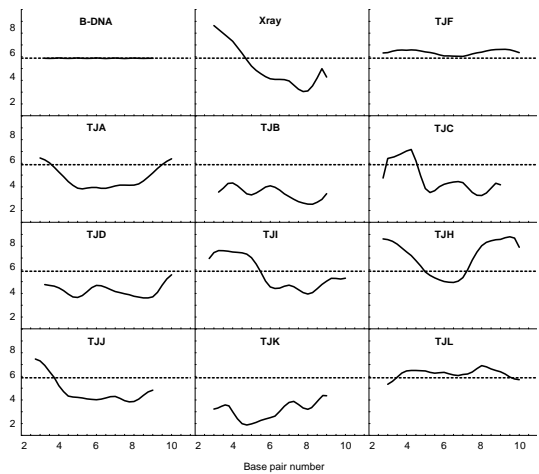,width=7cm,angle=0.,%
bbllx=120bp,bblly=200bp,bburx=480bp,bbury=570bp,clip=t}}
\caption{\label{Falmg}
The profiles of the minor grooves for the structures
shown in Fig. \protect\ref{Falstr}. The local minor groove width
is given in angstr{\"o}ms as computed with
the spline algorithm implemented in Curves.
\protect\cite{Stofer:94,Curves:} The dotted line indicates the
level corresponding to the canonical B-DNA.
}\end{figure}

Tables \ref{Ttjli}, \ref{Thlpa} and Figs. \ref{Falstr} and \ref{Falmg}
characterize in general the numerical experiments and results
considered in detail in sections below. Table \ref{Ttjli} summarizes
variations of computational protocols for different trajectories. For
brevity each trajectory is referred to by using a single-letter code
with a prefix ``TJ''. Trajectories from A to K computed for the
{\em Eco}RI dodecamer differ either by the starting state, or by the
details in the application of hydration procedures, or by the reduced
phosphate charge. TJA, TJB, and TJC have been the subject of our
previous report. \cite{Mzjacs:98} They are discussed together with the
rest and are included in Table \ref{Ttjli} for completeness. It is
worth reminding here that TJC was computed by using the conventional
Cartesian coordinate MD method starting from the same state as TJA.
TJD and TJI also started from the same conformation, namely, the Xray
structure, \cite{Drew:81} with
significantly different initial distribution of water
molecules obtained by varying parameters of the hydration algorithm.
TJE, TJF, and TJG were computed in the dehydrated state of the duplex
to clarify the role of the explicit water and study the effect of
phosphate screening. Similarly, TJH, TJI, TJJ, and TJK were calculated
for the same model system with different phosphate screening
parameters. Finally, TJL computed for another sequence checks the
possibility of observing sequence dependent structural effects with
this DNA model.

\begin{table}[h]\caption{\label{Thlpa} Selected Helical Parameters of
Different DNA Conformations}
\begin{tabular}[t]{|dddddd|}
         & Xdisp & Inclin & Propel & Rise & Twist\\\hline
A\tablenotemark[1]
         & -5.43 &  19.12 &  13.70 & 2.56 & 32.70\\
B\tablenotemark[1]
         & -0.70 &  -5.98 &   3.87 & 3.38 & 36.01\\
Xray\tablenotemark[2]
         & -0.54 &   0.19 & -13.45 & 3.37 & 35.89\\
TJA      & -1.10 &   2.06 & -12.54 & 3.31 & 35.67\\
TJB      & -0.95 &   0.62 & -12.80 & 3.37 & 36.36\\
TJC      & -1.40 &   4.04 & -16.30 & 3.36 & 34.69\\
TJD      & -1.41 &   1.83 & -12.55 & 3.34 & 35.82\\
TJE      & -3.28 &   2.98 &  -5.62 & 3.41 & 31.13\\
TJF      & -3.14 &   2.66 &  -5.16 & 3.44 & 31.65\\
TJG      & -3.08 &   1.73 &  -3.96 & 3.49 & 31.79\\
TJH      & -2.05 &   3.94 &  -7.39 & 3.40 & 32.08\\
TJI      & -1.65 &   2.43 & -10.21 & 3.39 & 34.15\\
TJJ      & -1.24 &   -.32 & -11.50 & 3.37 & 34.84\\
TJK      & -0.17 & -11.91 & -10.54 & 3.51 & 34.89\\
TJL      & -0.37 &  -2.24 &  -4.53 & 3.42 & 34.56\\
\end{tabular}
\tablenotetext[1]{
Canonical DNA forms \cite{Arnott:72} constructed with
NUCGEN procedure of AMBER \cite{AMBER:} for the {\em Eco}RI dodecamer. }
\tablenotetext[2]{
The crystal conformation of the {\em Eco}RI dodecamer. \cite{Drew:81}
(file 1bna in Protein Database. \cite{PDB:}) }
\end{table}

Figure \ref{Falstr} presents the final states of representative
trajectories in Table \ref{Ttjli} together with the canonical B-form
and the Xray structure of the {\em Eco}RI dodecamer. The drawings are the
outputs of Curves program \cite{Curves:} which also computes the
optimal helical axis shown by a continuous roughly vertical line in
each plate. Figure \ref{Falmg} shows variations of the minor groove
widths for the same structures computed by the spline algorithm
implemented in Curves. \cite{Stofer:94} On the right of each
structure in Fig. \ref{Falstr} the rmsd from the canonical B-DNA is
given. For {\em Eco}RI dodecamer the second value shows the rmsd from
the Xray conformation. This structure is asymmetrical and, since the
sequence is palindromic, the rmsd can be computed in two orientations.
Both orientations were tried and the lower rmsd values are given in
Fig. \ref{Falstr}. It is seen that the structures are generally close
to the two experimental B-DNA conformations, with a few exceptions to
be discussed hereafter. The majority of the computed structures of the
{\em Eco}RI dodecamer have a characteristic profile of the minor groove with
a narrowing in the center, which is a well-known feature of the
experimental crystallographic structure. \cite{Kopka:83}
These results confirm our
initial report on these simulation conditions. \cite{Mzjacs:98} Time
dependencies of rmsd's are not analyzed here systematically. We note
only that for trajectories with the final rmsd values below 2~\AA\
they were qualitatively similar to those reported for TJA, TJB and
TJC. Namely, after the first 100 ps the rmsd from
the canonical B-form and the Xray structure are usually found between
2.5 and 3.5~\AA. In our earlier calculations \cite{Mzjacs:98} this
large initial shift was either already present in the initial
conformation or resulted from {\em in vacuo} energy minimization of
the canonical B-DNA form, but it also occurs in trajectories
starting from the Xray conformation. During the first two nanoseconds
the rmsd values gradually decrease, with the 2~\AA\ level commonly
passed within the second nanosecond. The lowest rmsd values from
experimental B-DNA conformations are usually reached within the first
three nanoseconds and next the drift becomes below the level of
fluctuations in the nanosecond time scale.

Selected averaged helical parameters corresponding to the structures
in Fig. \ref{Falstr} are assembled in Table \ref{Thlpa}. The complete
set of these parameters have been discussed for TJA, TJB and TJC,
\cite{Mzjacs:98} and here we note only that parameters not included in
Table \ref{Thlpa} do not show systematic deviations from experimental
values. In contrast, Xdisp and inclination obtained in different
simulations for {\em Eco}RI sequence show a small but systematic deviation.
Propeller is generally well reproduced, and it is included in Table
\ref{Falstr} because it plays an important structural role for AT-rich
sequences \cite{Coll:87} and is rather sensitive
to simulation
conditions. Finally, twist and rise are the key helical parameters
responsible for the major part of rmsd's. The time variation of the
helical parameters in the present B-DNA model has already been
characterized. \cite{Mzjacs:98} It is similar to the above described
behavior of rmsd's in the sense that it usually takes about 3 ns for
the main parameters to come close to the standard B-DNA values. In all
calculations reported here the initial large rmsd's are accompanied by
low average twist which usually raise from 32--33\degree to
34--35\degree during calculations.

The data presented in Figs. \ref{Falstr} and \ref{Falmg}, and Tables
\ref{Ttjli} and \ref{Thlpa} are discussed in detail in sections below.

\subsection{B-DNA Dynamics in the Dehydrated State}
It is known since long ago that reasonably good duplex DNA
conformations can be obtained by minimization of potential energy with
implicit modeling of solvent effects, notably, by reducing phosphate
charges and scaling of Coulomb forces with a distance dependent
electrostatic function. \cite{Lavery:94} On the other hand, it is also
known that these structures are not dynamically stable. \cite{Swaminathan:91}
The dehydrated state presents non-negligible theoretical interest because
the ``naked DNA'' is the core of the whole system and understanding
its mechanics is important for distinguishing between the trends that
originate from intra-duplex interactions and water effects. We briefly
report here about a few interesting lessons that we have learned from
ICMD simulations of B-DNA in the dehydrated state.

We hoped initially that, due to the reduced flexibility of the
standard geometry DNA model, its structure could be more stable than
in the earlier Cartesian coordinate MD simulations. It was found,
however, that trajectories generally tend to be trapped in earlier
described deformed states characterized by a collapsed minor groove.
\cite{Swaminathan:91} Such transformations invariably start from a phosphate $\rm
B_I\to B_{II}$ flip, which can next result in a significant deformation
of the standard B-DNA shape, with partial local closing of the minor
groove. This behavior seems to be an intrinsic defect of {\em in
vacuo} conditions. However, the probability of collapsing transitions
was high only during the first nanosecond when, in animation, the DNA
molecule exhibited very violent periodical low frequency bending and
stretching motions. This intermediate phase is caused by the heating
procedures that commonly last less then 10 ps and always use some
velocity scaling which, however, affects only normal modes that have
non-zero instantaneous velocities. The slowest normal modes of a
dodecamer B-DNA fragment have periods beyond 10 ps, \cite{Lin:97}
therefore, during
heating, those that happen to be in their fast phase are strongly
overheated. {\em In vacuo} the uneven distribution of the kinetic
energy is rather persistent and it apparently takes hundreds of
picoseconds to reach equipartition. In many test calculations we
observed that if the structure managed to survive the first
nanosecond, it continued to stay in a reasonably good B-DNA form for
rather long time so that its average dynamic properties could well be
characterized. $\rm B_I\to B_{II}$ transitions still occur, but the
minor groove does not close immediately and many such transitions
appear reversible.

Trajectories TJE, TJF, and TJG in Table \ref{Ttjli} have been computed
with special precautions that take into account the foregoing
qualitative features. Dynamics were initiated with random velocities
and during the first nanosecond the temperature was raised to 250~K by
the Berendsen algorithm \cite{Berendsen:84} with a relaxation time of
100~ps. Calculations were next continued with the standard
parameters, that is with the bath temperature of 300~K and a 10~ps
relaxation time. If a transition to a collapsed state was observed
the calculations were restarted from one of the preceding
conformations with random velocities at 300~K. Such restarts were
repeated until a sufficiently long continuous trajectory have been
obtained. In the case of TJF one restart was necessary to get a
continuous 5~ns trajectory. TJG was also stable during 5~ns after one
restart, but only the first 2.9 ns were used for analysis because a
few reversible $\rm B_I\to B_{II}$ transitions that happened in its
second part caused non-negligible temporary distortions. In contrast,
in the case of TJG three restarts were necessary to obtain a 2.7~ns
continuous run at the end of which the structure collapsed.

\begin{figure}
\centerline{\psfig{figure=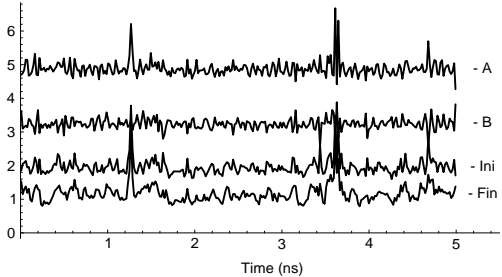,width=7cm,angle=0.,%
bbllx=180bp,bblly=300bp,bburx=430bp,bbury=470bp,clip=t}}
\caption{\label{Fnowt}
The time variations of rmsd from several
reference structure for TJF. A, B, Ini and Fin correspond to
the canonical A and B forms, the initial minimum energy structure,
and the final MD state, respectively. All rmsd values are 
given in angstr{\"o}ms.
}\end{figure}

Figure \ref{Fnowt} shows the time variations of rmsd from several
reference structure for TJF. As wee see, the {\em in vacuo} dynamics are
qualitatively different from that with explicit hydration. Namely, the
short-time rmsd fluctuations are similar or even larger,
but the trajectory apparently samples from a broad energy
valley around one and the same state in which it appeared
after equilibration. This system has an
essentially constant list of atom-atom contacts.
The local energy minima can occur due to complex energy profiles of
base stacking, but they seem to be rather shallow and the trajectory
passes above the barriers so that no distinct conformational
transitions within the B-DNA family are observed.
A few sharp peaks in the traces in Fig. \ref{Fnowt} correspond
to short-living $\rm B_I\to B_{II}$ transitions. The rmsd between
the structures averaged over the five consecutive nanosecond intervals
was as low as 0.1~\AA. The final TJF state shown in Fig. \ref{Falstr}
has rmsd of only 0.5~\AA\ from the initial minimum energy
structure.

Thus, a naked B-DNA duplex presents a relatively simple object
with a single significant energy minimum which is rather broad.
This state resembles the canonical B-DNA and is characterized
by the rmsd's of 3.1 and 4.5~\AA, respectively, from B and A forms
and some of its helical parameters shown in Table \ref{Thlpa}
are also between the canonical A and B-DNA values.
The other two {\em in vacuo} trajectories were very similar.
This means, in particular, that the shorter durations of TJE
and TJG are quite sufficient for correct averaging.
A visual analysis of the final TJF state superposed with the canonical
B-DNA shows that the relatively high rmsd
value results from the following main contributions: (i) lower
negative Xdisp, (ii) much lower average twist, and (iii) small but
non-negligible bending of the helical axis towards the center of the
minor groove. The latter effect is very clear in superposition figures
and it is nicely detected by the Curves algorithm. \cite{Curves:} The
first two features give this structure some A-like character although
one notes that in fact only these two helical parameters clearly
deviate towards the A form. There was only very minor difference
between the final states of these three {\em in vacuo} trajectories,
which is discussed hereafter.

\subsection{Negative Role of Water Films in Major Groove}

\begin{figure}
\centerline{\psfig{figure=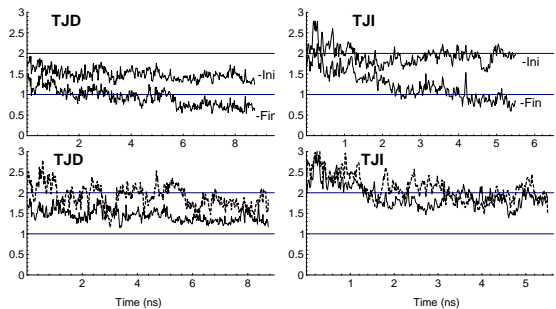,width=7.5cm,angle=0.,%
bbllx=130bp,bblly=280bp,bburx=490bp,bbury=500bp,clip=t}}
\caption{\label{FtjX}
Comparison of the time variations of rmsd from reference
conformations for TJD and TJI. Ini and Fin mark the traces of
rmsd from the initial energy minimum and the final
MD state, respectively. The lower plates exhibit variation
of the ``proper'' and ``improper'' rmsd's from the crystal
structure \protect\cite{Drew:81} shown by the solid and broken
lines, respectively. All rmsd values are 
given in angstr{\"o}ms.
}\end{figure}

The hydration procedure described above has been designed to
completely fill up the minor groove of B-DNA and also left in the
major groove some small number of water molecules to which we
initially did not pay much attention. The effect of
this water is not negligible, however. Figure \ref{FtjX} compares the
time variations of rmsd from reference conformations for TJD and TJI.
In both cases the initial state of was prepared by hydrating the
crystallographic structure which, because of the water crowding, has
only slightly changed during energy minimization (the rmsd 1.25~\AA). In the
crystal conformation, the two opposite phosphate traces come close to
each other in the middle, which is usually described as narrowing of
the minor groove. \cite{Kopka:83} For this specific shape our
hydration procedure left in the major groove a relatively large
number of water molecules.
The left two plates in Fig. \ref{FtjX} show that, in dynamics, the
structure slowly returns to the initial state after first leaving it
rapidly during heating. The lower plate shows the time variation of
the rmsd from the crystal structure. Due to the symmetry of the
sequence the rmsd comparison can be made in two ways giving the
``proper'' and ``improper'' rmsd values, the latter corresponding to
the opposite assignment of the two strands. The crystal conformation
itself is strongly asymmetric and characterized by a large
``improper'' rmsd of 1.7~\AA. In dynamics, this initial asymmetry
should relax, and the same level must be reached by the two rmsd
values after an ultimate equilibration. Figure \ref{FtjX} shows that
although the TJD trace of the ``proper'' rmsd reaches very low values
it is systematically shifted downward from the ``improper'' trace. In
other words, during these nine nanoseconds this symmetric DNA duplex
always ``remembers'' the asymmetry of its initial state.

The right two plates in Fig. \ref{FtjX} exhibits analogous traces for
TJI where all major groove water molecules were removed at the
beginning. The total number of water molecules appears larger
due to small variation of the hydration procedure, notably,
the initial 5~\AA\ shell was constructed with all crystallographic
waters considered as part of the duplex. Note that the structure again
moves out from the initial energy minimum and back, but the maximal
rmsd value reaches 2.8~\AA\, which is much larger than in the previous
case. The deviation from the crystal conformation is also one angstrom
larger, and, importantly, the two rmsd traces level after 3 ns
indicating that the ``memory'' of the initial asymmetric conformation
is essentially lost.

This example illustrates the general effect observed in many other
simulations. When the number of water molecules is not sufficient to
form a competent shell they cannot diffuse and tend to occupy fixed
positions thus stabilizing some random conformations and slowing down
the sampling. Very often they form films that connect several bases
and phosphate groups and can induce local bending of the helical axis.
To reduce these effects, in the last few trajectories in Table
\ref{Ttjli} the major grove was cleaned from the remaining water
molecules before heating. This does not mean that we consider
interactions in the major groove unimportant. The present DNA model
essentially performs the structural analysis of water surrounded by
the mobile DNA walls in the minor groove. The relatively small number
of these molecules makes possible nearly exhaustive sampling. In doing
so we implicitly assume that, due to much larger number of solvent
molecules involved in hydration of the major groove, its water shell
can easily take any necessary shape corresponding to a given DNA conformation.
Our calculations show that an empty major groove is a much better
approximation of such behavior than a faintly hydrated one.

\subsection{Characteristic Hydration Patterns}

\begin{figure}
\centerline{\psfig{figure=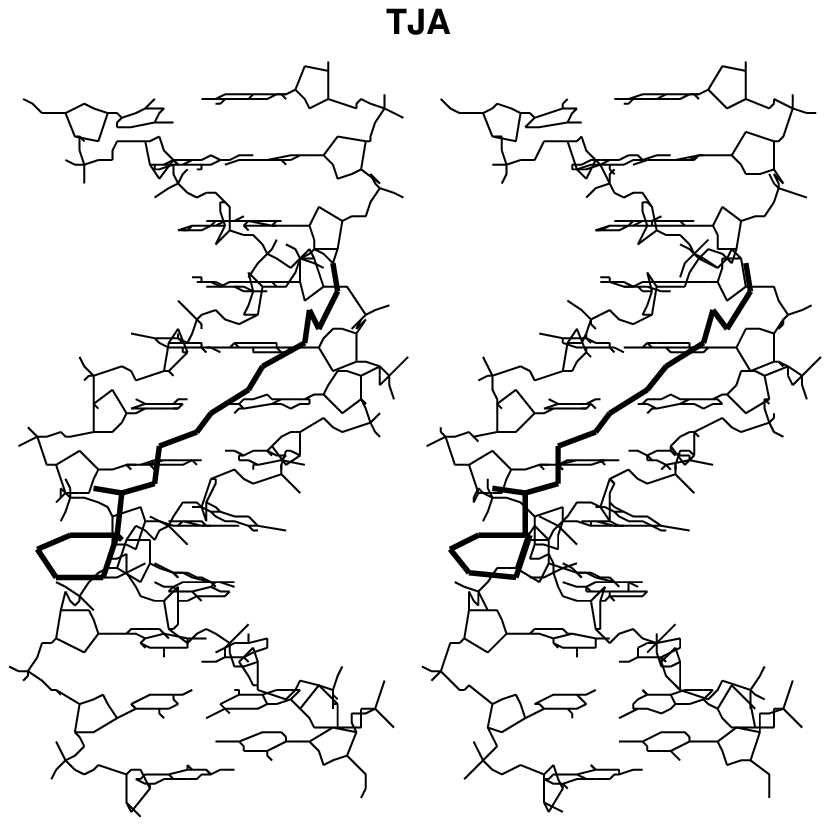,width=7.5cm,angle=0.,%
bbllx=150bp,bblly=200bp,bburx=430bp,bbury=450bp,clip=t}}
\centerline{\psfig{figure=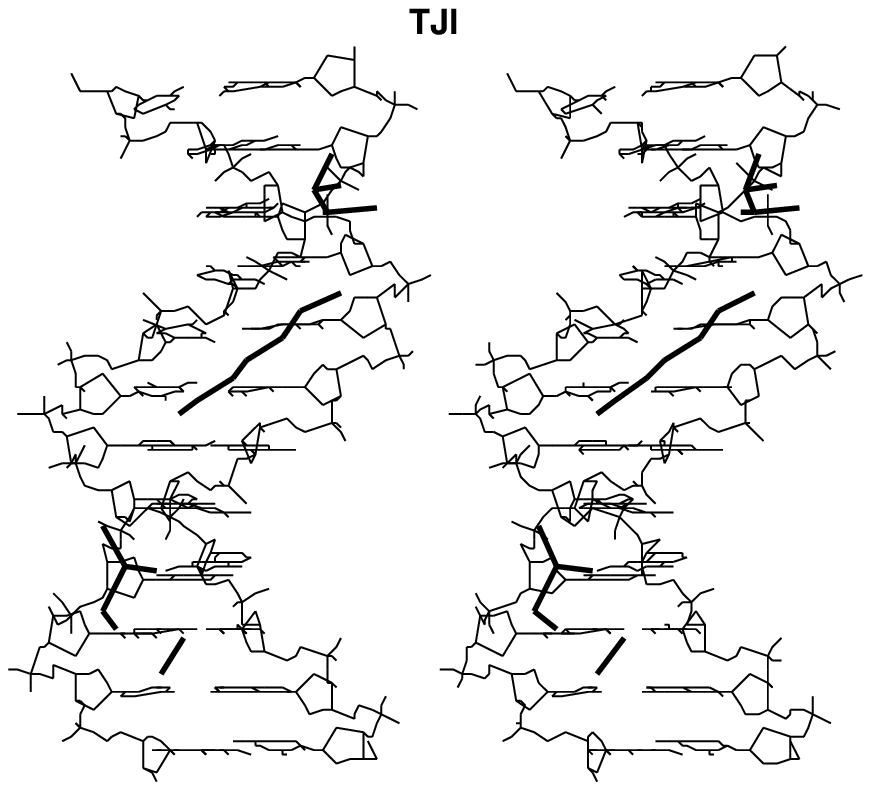,width=7.5cm,angle=0.,%
bbllx=150bp,bblly=280bp,bburx=430bp,bbury=530bp,clip=t}}
\caption{\label{Fstereo}
The final states of TJA and TJI with the most stable hydration
sites in the minor groove. The average oxygen positions
of 20 least mobile water molecules in the minor groove
are joined by the thick lines if they are close enough for
hydrogen bonding.
}\end{figure}

Figure \ref{Fstereo} shows the final states of TJA and TJI with the
most stable hydration sites in the minor groove. Atom coordinates from
the trajectory points saved during the last nanosecond were
superimposed, and the position fluctuations were computed for all
water molecules. Figure \ref{Fstereo} shows average oxygen positions
for 20 least mobile water molecules in the minor groove, with the
sites close enough for hydrogen bonding joined by the thick lines. The
amplitudes of the position fluctuations range from 0.5 to 1.0~\AA\ and
are close to that of the neighboring DNA atoms indicating a tight
binding.  In case of TJA similar sites are also found in the major
groove, but as we noted above they are not representative and we do
not discuss them here. In TJA the slowest water molecules form a
continuous cluster at the bottom of the minor groove. In contrast, in
TJI several such clusters are separated by areas where water is more
mobile. In both cases, however, in the middle of the minor groove a
linear trace is found corresponding to the characteristic hydration
pattern often observed in crystals and referred to as the ``hydration
spine''.  \cite{Dickerson:82}

\begin{figure}
\centerline{\psfig{figure=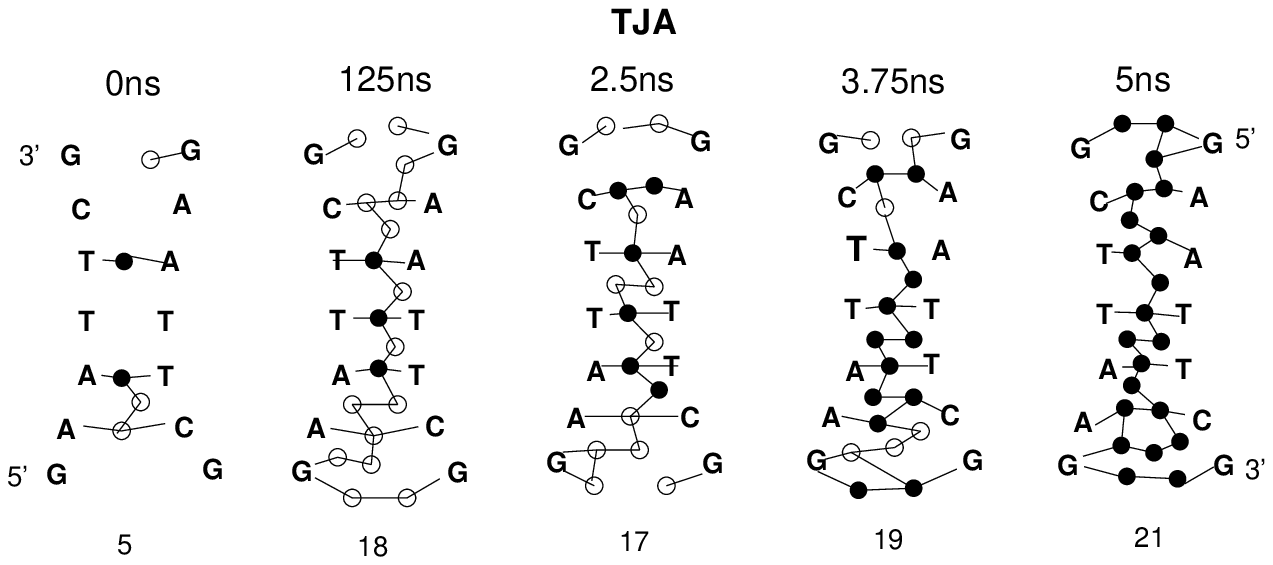,width=7.5cm,angle=0.,%
bbllx=10bp,bblly=10bp,bburx=375bp,bbury=175bp,clip=t}}
\centerline{\psfig{figure=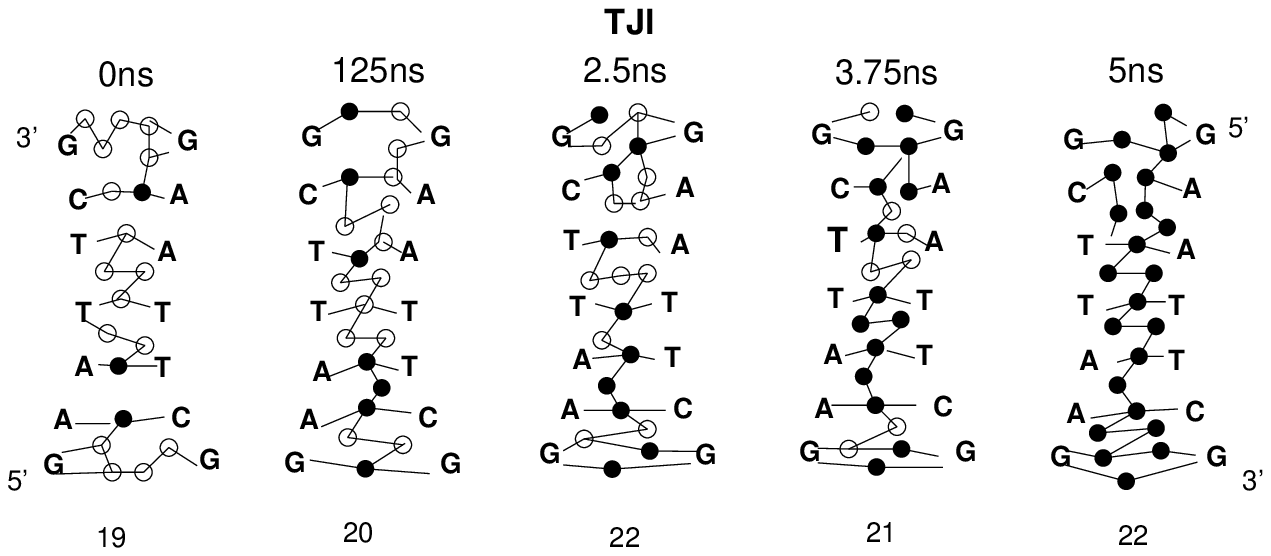,width=7.5cm,angle=0.,%
bbllx=10bp,bblly=10bp,bburx=375bp,bbury=175bp,clip=t}}
\caption{\label{Fspine}
Schematic representation of the time evolution of the
minor groove hydration patterns in TJA and TJI. The
hydrogen bonding contacts shown for the central eight base
pairs were taken from individual snapshots as explained in
the text. The filled circles denote water molecules that keep their
orientations in all subsequent snapshots.
}\end{figure}

Figure \ref{Fspine} shows schematically the time evolution of the
minor groove hydration patterns in these two trajectories. The
hydrogen bonding contacts shown for the central eight base
pairs were taken from individual snapshots as follows.
First, water molecules directly
bound to bases were considered and the shortest water bridges between
them were found. Only one and two member bridges are shown in the
figure. The filled circles denote water molecules that keep their
orientation in all subsequent snapshots. The orientation is
considered to be maintained if there is at least one chain to a
correct water-base contact. For clarity the opposite strands are
shifted by one step with respect to the Watson-Crick pairing.

In TJA most hydration sites were vacant after heating while already at
1.25 ns they are all occupied, with the canonical spine structure
formed in the middle which spans over six consecutive base pairs. The
total number of water molecules selected by the foregoing rules slowly
increases during the whole trajectory; the spine shortens and is
gradually replaced by less economical patterns near both ends. The
number of sites reaches 21 after 5 ns, but we note that the TJA
plate in Fig.  \ref{Fspine} actually shows 32 different water
molecules, which manifests a certain level of exchange with the
environment. It is readily seen that the three water molecules that
form the first layer of the hydration spine are the most stable in
this system. At the same time one notes that a transition from spine
to ribbon type of hydration \cite{Edwards:92} occurs at the upper
border of the spine.

The initial state for the TJI was prepared from the Xray structure
with new water molecules added to already occupied experimental
hydration cites. That is why in TJI the minor groove hydration is
rather dense already after heating. The total number of solvent
molecules selected by the chosen rules fluctuates at the level reached
in TJA after 5 ns. As a whole the evolution of hydration patterns is
rather different in TJI compared to TJA, but still water molecules
at the bottom of the hydration spine are among the least movable. Note
also that in this case a reversible transition from the spine to the
ribbon type of hydration occurs at the upper spine border.

Spontaneous formation of the hydration spine
in the middle of the minor groove has
been observed in all trajectories computed for the {\em Eco}RI dodecamer.
This characteristic water arrangement has been first encountered in
the crystal structure of this duplex \cite{Dickerson:82}
and is generally considered as a characteristic
feature of certain AT-rich sequences that tend to adopt an unusual
conformation with a particularly narrow minor groove. \cite{Dickerson:92} We
may conclude, therefore, that the present B-DNA model nicely
reproduces this well established experimental observation, which
is the most probable origin of a surprisingly good correspondence between the
computed and experimental DNA conformations. It can be
noted also that the water structure in the minor groove is rather
persistent and changes very slowly. As we discussed above, due to the
initial ``cleaning'' of the major groove in TJI the DNA molecule
lost the ``memory'' of the initial state much faster than in TJD.
Figure \ref{Fspine} demonstrates, however, that throughout the
trajectory the hydration patter is more spine-like in the lower part
of the minor groove and more ribbon-like in the upper part, the
feature inherited from the initial Xray structure where the spine
expands from the center two times farther towards the narrower end of
the minor groove. \cite{Drew:81a}

It should be noted, finally, that the spine may be also a model
dependent feature. We have noticed that, although trajectories always
start from a well minimized structure, the dynamics usually begins
with rapid unwinding and extending of the DNA molecule, which narrows
the minor groove and pushes much of the water out. This occurs
regardless of the sequence and the spine is normally the first
hydration pattern that sets up afterwards. The subsequent slow
diffusion of water molecules into the minor groove usually takes
several nanoseconds and often goes faster near the ends than in the
middle. Nevertheless, the high stability of this specific hydration
pattern in the center of the {\em Eco}RI dodecamer as well as for some other
sequences like A-tracts, which are not considered here, is a sequence
dependent feature. To support the last assertion we included in Tables
\ref{Ttjli} and \ref{Thlpa} and also in Figs. \ref{Falstr} and
\ref{Falmg} trajectory TJL computed in similar conditions for a
GC-alternating dodecamer. Table \ref{Thlpa} shows that, except for the
twist, the final TJL state has average helical parameters even closer
to the canonical B-DNA than {\em Eco}RI dodecamer. Notably, its inclination
is negative and Xdisp is much less than in any {\em Eco}RI trajectory. In
terms of rmsd it is similarly close to the canonical B-DNA. At the same
time, the hydration patterns in the minor groove were very different
from those discussed above and it is seen in Figs. \ref{Falstr} and
\ref{Falmg} that the final TJL state has the minor groove evenly wider
than in the canonical B-DNA form. The minor groove was narrow in the
initial phase of the trajectory and then gradually opened. The
distinct properties of regular sequences in this DNA model will be
discussed in detail elsewhere.

\subsection{Long Range Electrostatics}
The most surprising property of the present DNA model is that its
trajectories converge to conformations that much better compare with
experimental data than in more expensive calculations with explicit
counterions and rigorous treatment of electrostatic interactions.
\cite{Mzjacs:98} This would be less surprising if the adjustable
parameters of the simple {\em ad hoc} treatment employed here were
specifically fitted to experimental data, but, in fact, we have
arbitrarily chosen one of the options earlier used in conformational
analysis of DNA, not even the best recommended. \cite{Flatters:97}
These observations suggest that the precision of calculation of long
range electrostatic interactions may be generally less important for
DNA structure than it is sometimes supposed. A comprehensive analysis
of this fundamental issue is beyond the capabilities of this simple
approach, nevertheless, it is interesting to check how this B-DNA
model responds to variation of parameters involved in calculation of
electrostatic interactions.

The {\em in vacuo} trajectories TJE, TJF, and TJG were computed with
phosphate charges varied from -0.25 to -0.75. They produce very
similar structures with rmsd between the final TJE and TJG states
around 0.5~\AA\ and the final TJF state at roughly 0.3~\AA\ from the
both. Table \ref{Thlpa} exhibits very small, but regular variation of
helical parameters. Namely, growing phosphate repulsion moves Xdisp,
inclination and twist closer to the canonical B-DNA values, but
propeller and rise in the opposite direction. The rmsd's of the
final MD states from the reference experimental DNA conformations are
practically identical for all three trajectories.

The same range of phosphate charges is covered by trajectories TJH,
TJI, TJJ and TJK where the minor groove has been hydrated. Table
\ref{Thlpa} shows that in this case variations of helical parameters
are stronger and less regular, but some of them, notably, Xdisp,
inclination and twist exhibit the same trends as {\em in vacuo}.
Reduction of phosphate repulsion accentuates already existing small
deviations from canonical B-DNA values towards A-DNA. The increased
repulsion has an opposite effect and one may expect that, in terms of
helical parameters, the best fitting of the computed and experimental
conformations can be achieved with phosphate charges slightly below
-0.6 aeu. Unexpectedly, however, variation of phosphate charges in an
explicitly hydrated B-DNA model significantly affects the overall
bending of the molecule which is seen in Fig. \ref{Falstr}. In fact,
all computed conformations of the {\em Eco}RI dodecamer except TJB are
slightly bent towards the center of the minor groove. It appears that
the increase in the phosphate repulsion also increases this bending,
which is a non-trivial effect of the minor groove hydration since no
such trend is observed in the {\em in vacuo} conditions.
This effect deserves careful investigation because it in fact corresponds
well to certain experimental observations. Gel retardation experiments
with sequence repeats indicate that, in solution, the {\em Eco}RI
dodecamer should be bent towards the center of the minor groove.
\cite{Diekmann:88} It is also known that DNA bending is generally reduced in
high salt \cite{Diekmann:87} which also reduces the persistence length of
DNA, \cite{Hagerman:88} with the latter effect attributed to reduced
phosphate repulsion due to better screening. Comparison of the final
states of TJH, TJI, TJJ and TJK suggests that, in the minimal B-DNA,
bending is also somehow connected with phosphate repulsion
and increased backbone stiffness also favors bending.

Because of the foregoing effect the rmsd's from experimental B-DNA
conformations does not follow the corresponding deviations
of helical parameters. In the case of TJJ, for instance,
the increased bending overshadows the small improvement in helical
parameters with respect to the canonical B-DNA.
These results generally show, however, that significant variations of the most
arbitrary parameter involved in our simplistic treatment of the
electrostatic effects still result in structures that are close
to experimental B-DNA both in terms of helical parameters and atom rmsd's.
Note that the largest rmsd in Fig.
\ref{Falstr} observed for TJH is lower than in all reported
simulations with the particle-mesh Ewald method. \cite{Mzjacs:98}
Although its helical parameters
noticeably deviate from canonical values it still qualitatively
reproduces the characteristic shape of the minor groove with a short
hydration spine in the center.

\subsection{Concluding Discussion}
The minimal model of B-DNA explored here consists of two qualitatively
different components, namely, the DNA duplex
and a water cloud docked in the minor groove.
Dynamics simulations of a naked B-DNA duplex show that it presents a
relatively simple object with a broad energy valley around a single
state with an intermediate conformation between A and B DNA forms.
This state is ``fragile'' in the sense that $\rm B_I\to
B_{II}$ transitions, which are generally believed to introduce only
subtle nuances in the B-DNA structure, \cite{Berman:97} cause gross
structural transitions to an essentially different DNA form with a
collapsed minor groove. \cite{Swaminathan:91} With a water cloud docked in
the minor groove the system acquires qualitatively different properties.
Its potential energy landscape becomes rough, with multiple
significant local minima corresponding to
rearrangement of atom-atom contacts in water and at
the DNA-water interface. Dynamics simulations of this system can be
viewed as structural analysis of water near a mobile surface that
carries a net negative charge and has a specific distribution of
hydrogen bonding contacts.

The main advantage of these simulation conditions is the
possibility of practically exhaustive sampling with the presently
available computer resources. It should be noted that the
sampling and the affordable duration of trajectories are not strict
synonyms. The number of essential degrees of freedom is another
important parameter that should not be ignored. We expect, for
instance, that, in otherwise similar conditions, the rate of
convergence would be lower for a B-DNA duplexes with both minor and
major grooves hydrated explicitly. That is why, in our opinion, the
minimal B-DNA should be distinguished and used, notably,
for equilibration of the surface water in preparative phases of
more expensive DNA simulations.

The properties of this system are non-trivial and sometimes
counter-intuitive. We have shown here that the canonical B-DNA form
presents a strong point of attractions in its conformational space, so
that numerous trajectories of the {\em Eco}RI dodecamer and one long
trajectory of the GC-alternating dodecamer converge to conformations
with nearly canonical helical parameters. We have shown that the
sequence specific features of the {\em Eco}RI dodecamer are nicely
reproduced, namely, the profile of the minor groove and spontaneous
formation of the characteristic ``spine'' hydration pattern
which is generally considered as an indispensable part of this
structure. \cite{Dickerson:82} To our knowledge, none of the numerous earlier
theoretical studies of this molecule could reach a comparable
level of agreement with experimental data.
All these observations justify the minimal model of B-DNA, confirm its
credibility and suggest that it has a potentially wide scope of application.

\appendix
\section{Corrected Implicit Leapfrog Integrator}
The quasi-Hamiltonian equations of motion used in the ICMD method have
the following general form \cite{Mzjcc:97}

{\mathletters\label{EqH}\begin{eqnarray}
{\bf\dot p}&=&
{\bf f}({\bf q})+{\bf w}({\bf q},{\bf\dot q})\\
{\bf\dot q}&=&{\bf M}^{-1}{\bf p}
\end{eqnarray}}
where the dot notation is used for time derivatives. Vectors ${\bf
q}$, ${\bf p}$, and ${\bf f}$ denote generalized coordinates,
conjugate momenta, and generalized forces, respectively, ${\bf M}$ is
the mass matrix and ${\bf w}({\bf q},{\bf\dot q})$ is the inertial
term. The implicit leapfrog integrator used for systems with flexible
internal rings reads \cite{Mzprep:98,Mzjchp:99}
{\mathletters\label{Elfcn}
\begin{eqnarray}
{\bf f}_n&=&{\bf f}({\bf q}_n)\\
\circ\ {\bf q}_{n+\frac 12}&=&{\bf q}_n+{\bf\dot q}_{n+\frac 12}\frac h2\\
\nonumber\circ\ {\bf\tilde p}_{n+\frac 12}&=&{\bf p}_{n-\frac 12}+{\bf f}_nh+\\
&&\left({\bf w}_{n-\frac 12}+{\bf w}_{n+\frac 12}\right)\frac h2+{\bf
f}^\bot_{n-\frac 12}\frac h2\\
\circ\ {\bf\dot q}_{n+\frac 12}&=&{\bf T}_{n+\frac 12}{\bf
M}^{-1}_{n+\frac 12}
{\bf\tilde p}_{n+\frac 12}\\
{\bf p}_{n+\frac 12}&=&{\bf M}_{n+\frac 12}{\bf\dot q}_{n+\frac 12}\\
{\bf f}^\bot_{n-\frac 12}\frac h2&=&{\bf p}_{n+\frac 12}-{\bf\tilde
p}_{n+\frac 12}\\
{\bf q}_{n+1}&=&{\bf q}_n+{\bf\dot q}_{n+\frac 12}h
\end{eqnarray}}
where the conventional notation is used for denoting on-step and
half-step values. Vector ${\bf f}^\bot$ denotes additional generalized
forces that originate from ring closure constraints. The corresponding
term is evaluated in Eq. (\ref{Elfcn}f) and presents just an
intermediate internal variable. Matrix ${\bf T}$ denotes a projection
operator. \cite{Mzprep:98,Mzjchp:99} The lines marked by circles are iterated
until convergence of Eqs.  (\ref{Elfcn}b) and (\ref{Elfcn}c). For tree
topologies this integrator is reduced to \cite{Mzjcc:97}

{\mathletters\label{Elf}
\begin{eqnarray}
{\bf f}_n&=&{\bf f}({\bf q}_n)\\
\circ\ {\bf q}_{n+\frac 12}&=&{\bf q}_{n-\frac 12}+
\left({\bf\dot q}_{n-\frac 12}+{\bf\dot q}_{n+\frac 12}\right)\frac
h2\\
\circ\ {\bf p}_{n+\frac 12}&=&{\bf p}_{n-\frac 12}+{\bf
f}_nh+
\left({\bf w}_{n-\frac 12}+{\bf w}_{n+\frac 12}\right)\frac h2\\
\circ\ {\bf\dot q}_{n+\frac 12}&=&{\bf M}^{-1}_{n+\frac 12}
{\bf p}_{n+\frac 12}\\
{\bf q}_{n+1}&=&{\bf q}_n+{\bf\dot q}_{n+\frac 12}h
\end{eqnarray}}

These integrators had been in use for some time when an important
failure has been revealed. Note that, unlike the classical leapfrog,
algorithm Eqs. (\ref{Elf}) computes both on-step and half-step
coordinates, even though the forces need not be computed for half
steps. The two sets of coordinates are coupled implicitly via the
iterative cycle Eqs. (\ref{Elf}b-\ref{Elf}d) where ${\bf f}_n$ and
${\bf w}_{n+\frac 12}$ depend upon on-step and half-step coordinates,
respectively. However, for each molecule the set of the generalized
coordinates includes three Cartesian coordinates of the first atom
\cite{Mzjcc:97} for which the inertial term in Eq. (\ref{EqH}) is zero
and thus the algorithm is reduced to the standard leapfrog. This means
that the on-step and half-step translations are uncoupled and the
corresponding half-step generalized coordinates can diverge.
They, however, affect angular velocities via Eq. (\ref{Elf}d) thus
becoming a hidden source of instabilities.

\begin{figure}
\centerline{\psfig{figure=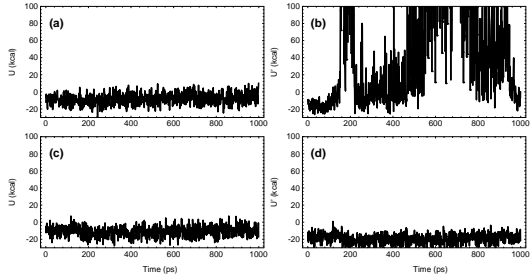,width=7.5cm,angle=0.,%
bbllx=70bp,bblly=250bp,bburx=550bp,bbury=520bp,clip=t}}
\caption{\label{Falg}
The time variations of the potential energy computed for on-step (U)
and half-step (U') coordinates, respectively, during a nanosecond
trajectory computed with the original algorithm (a,b) and the
corrected one (c,d).  }\end{figure}

The above effect is exposed in Fig. \ref{Falg}a and \ref{Falg}b which show
the time fluctuations of the potential energy computed for on-step and
half-step coordinates, respectively, during a nanosecond trajectory of
a hexamer DNA duplex with the sequence $\rm A_6\cdot T_6$. The model
system includes only two strands with no explicit water. During the
first 100 ps no suspicious symptoms can be detected, but after that the
half-step energies exhibit irregular fluctuations. They result from
atom clashes which do not affect the trajectory, and just serve here
as indicators of divergence. Figure \ref{Falg} also explains why this
problem has not been noticed immediately. Note that, in spite of the
apparent difficulties, the trajectory has been successfully finished
and we can add that the computed total energy was well conserved. For
a single polymer chain this defect is indistinguishable regardless of
the chain lengths. For ensembles of small molecules, like water, it
produces a small increase in the drift of the total energy after many
hundreds of picoseconds, which does not appear, however, if the
trajectory is restarted periodically. It becomes significant only for
DNA duplexes of 15 base pairs and longer, but in this case the
instability develops too fast, without intermediate phase like in
Fig. \ref{Falg}b, and results in sudden crashes of calculations which
have been initially attributed to some unclear physical effects.

The problem is overcome by simply replacing Eqs. (\ref{Elf}b) and
(\ref{Elfcn}b) with

\beq\label{Emf}
{\bf q}_{n+\frac 12}={\bf q}_n+{\bf\dot q}_{n+\frac 12}\frac h2
\eeq
This equation explicitly couples the on-step and half-step coordinates
in the iteration cycle, which immediately eliminates all artifacts
described above. Figures \ref{Falg}c and \ref{Falg}d demonstrate the
performance of the refined integrator (\ref{Elfcn}) for a trajectory
starting from the same state as in Figs. \ref{Falg}a and \ref{Falg}b. It
is not difficult to verify that Eq. (\ref{Emf}) does not change
neither the order nor the time reversibility of integrators
(\ref{Elf}) and (\ref{Elfcn}). It is not surprising, therefore, that
in standard quality tests, which normally employ relatively short
trajectories, this modification is absolutely neutral.

\section{Anonymous Comments}
This section contains comments from anonymous referees of two peer-review
journals where the manuscript has been considered for publication, but
rejected.
\subsection{Journal of Physical Chemistry}
\subsubsection {First referee}
{\bf Summary:}

In this paper, the author reports on results obtained from dynamics
calculations using a relatively new, ``minimal'' model for DNA salvation.
This model consists essentially of hydrating the minor groove of the
DNA and reducing the charges on the phosphate groups. Dynamics were
carried out where only torsions and bond angles on sugars were free to
change. Thus, this dynamics model consists of many fewer degrees of
freedom than the typical ``good quality'' DNA model, and dynamics can be
carried out with much less CPU time. A comparison of the results
using the minimal model indicates that these results are in very good
agreement with physical observation.

{\bf Review:}

I am, quite frankly, {\em very} surprised at the results the author
obtains with this minimal model. It is a widely-held belief that good
realistic results for DNA dynamics can only be obtained with an explicit
(periodic box) water model. More recently, it has been demonstrated that
with standard molecular dynamics and explicit periodic water, one must
also perform an Ewald type of calculations, whereby an infinite water
lattice about the DNA is simulated. (See, in particular, the work of the
Kollman and Darden groups). Certainly, the reduced phosphate model has
been held in some disdain for years now as a semi-physical kludge.

That said, proof is the results, and I (and everyone else working in the
DNA simulations field) will be very grateful if the minimal model
holds up to additional testing.

And therein lies the rub... For while the tests performed in the current
paper are enticing, they are not really complete enough to punch the
conclusions home. In particular, a direct head-to-head comparison for
the same DNA sequence to the results obtained an explicit (periodic
box) water model and Ewald summation is missing. I understand that such
an explicit water simulation is going to be much more cpu intensive.
But it would considerably strengthen the conclusions of the paper
to have such a comparison. In what ways do the results of the two
methods agree and disagree? Without knowing that, we can't really say,
concertedly, whether the minimal model produces crudely acceptable results
or something quantitatively better.

The other issue that worries me is alluded to in the paper. For the
sequence used for the bulk of the simulations in this paper, it is
well-known that the waters form a well-ordered ``spine of hydration''
in the minor groove. Could it be that the reason this minimal model
works so well is that {\em in this particular case} is that the
conformational preferences of the model are dominated by the spine of
hydration? If so, it might be that the minimal model will fail for
many other sequences. It would be reassuring to see the results of
a number of other sequences, including some that are not expected
to form a well-defined spine of hydration. The author does include
results for one sequence, poly GC, as an indication that good results
can be obtained even for sequences not generally observed to for the
well-defined spine of hydration. But additional non-AT rich examples
would be valuable.

Overall, I think the results presented in this paper are intriguing and
food for additional thought. To make this paper considerably stronger,
I would really like to see the comparison to explicit water dynamics.
But with that issue addressed (as well the point about different
sequences), this is definitely work I would like to see in print.

A final note: I am not fully convinced that J. Phy. Chem. would be
the best place for this work. The emphasis here is really on the
empirical biophysically-relevant results obtained with a previously
published model. There is relatively little discussion of physics
of the model itself. This is obviously and editorial judgment call,
however.

\subsubsection{Second referee}
This MS describes calculations attempting to model DNA with a partial
solvent model in which the minor groove is filled with explicit water
and the remainder of the environment of the molecule is treated with a
distance dependent dielectric screening function. Molecular dynamics are 
performed to determine the trajectory of the molecule in phase space
and compared with other relevant results.

In my consideration of this work, I cant really get beyond the problem
of why in the world anyone would want to work on a model like this.
Reports of MDs on DNA with fully explicit solvent including ions at
various concentrations as well are typical of recent works and projects
described in the current literature from the Kollman group, Pettitt
group and others. Thus it is not necessary to make this approximation.
Using this reduced model would end up having to be validated individually for each new sequence studied and there are surely cases it would fail. Thus
it is misdirected effort to proceed in this manner in my opinion, since
it is highly unlikely that this approach  is the basis for methodology
 anyone would want to seriously peruse. The fact that reasonable results
are obtained for the demonstration case does not ally my concerns.   

\subsection{Biopolymers}
\subsubsection {First referee}
The paper by Mazur is interesting and can be published in Biopolymers
as is, but it would be appropriate to have the author comment on the 
applicability of this model. Briefly, he describes how a minimal
model of DNA with only minor groove waters is stable under molecular
dynamics for nanoseconds and stays closer to expt than full solvation
model with counterions (he scales down the phosphate charges). In
this manuscript he shows how this model is effected by phosphate
charge and presence of major groove waters or exclusion of any
waters. These are interesting issues. The point that I would like
to have a comment is can such a model be sued for anything beside
duplex DNA in the B form. It appears to be not appropriate for simulating
the A to B transition because of the need to have changes in hydration of
the grooves during this process.

\subsubsection {Second referee}
There are some serious problems with this manuscript in both scholarship
and technique. As such I can not recommend it for publication. To be
specific:

1) Technically using the r-dielectric and explicit water is double
counting. Either the waters to close proximity are counted in the
explicit region or they are implicit. Allowing the r-dielectric in the
explicit region is explicitly overcounting the hydration in that region.
To add to that the phosphate charge reduction compound the physical
picture nears to the point of having a non-physical model with,
therefore, questional predictive power.

2) There is a deep problem with controls in this study. The amber force
field has an A-state which is not accessible without the direct influence
of coordinating ligands. Thus the B-states are all that left at ambient
temperature and salt conditions. This has been explored rather intensely
by a series of papers from the Kollman and Pettitt groups. The conclusion
left the paper makes is thus that DNA in this force field is not stable
with the r-dielectric, which is an old result, well known for some time.

\subsubsection {An adjudicator}
I have a lot of trouble seeing what sort of future is for the model discussed
here. At best, it applies only to B-form DNA with a particular force field,
and involves some rather arbitrary elements (concerning phosphate charges,
and which waters to include). The author's argument that it could be used
for ``exhaustive'' conformational searches does not seem very convincing
to me: if no significant sequence-dependent effects were to be found,
it would be of little interest, and if there were interesting structure
variations, I find it hard to believe that many readers would trust
the model used here, and calculations would need to be re-done anyway
with more complete representations of the solvent and counterion
environment. Given that the basic ideas have already been reported
(ref. 6), I would lean against publication of the manuscript in its
present form, which is largely devoted to comparisons to other
vacuum-like models whose limitations are already widely appreciated.

\end{document}